# C-MP: A decentralized adaptive-coordinated traffic signal control using the Max Pressure framework

Tanveer Ahmed[a,*], Hao Liu[a], Vikash V. Gayah[a]

*The Pennsylvania State University, University Park, PA, United States*

## Abstract

Coordinated traffic signals seek to provide uninterrupted flow through a series of closely spaced intersections, typically using pre-defined fixed signal timings and offsets. Adaptive traffic signals dynamically change signal timings based on observed traffic conditions in a way that might disrupt coordinated movements, particularly when these decisions are made independently at each intersection. To alleviate this issue, this paper introduces a novel Max Pressure-based traffic signal framework that can provide coordination even under decentralized decision-making. The proposed Coordinated Max Pressure (C-MP) algorithm uses the space mean speeds of vehicles to explicitly detect freely flowing platoons of vehicles and prioritizes their movement along a corridor. Specifically, upstream platoons are detected and their weight in the MP framework increased to provide priority, while downstream platoons are detected and their weight reduced to ensure smooth traffic flow across corridors. The study analytically proves that C-MP maintains the desirable maximum stability property, while micro-simulation analyses conducted on an arterial network demonstrate its ability to achieve a larger stable region compared to benchmark MP control policies. Simulation results also reveal that the proposed control algorithm can effectively coordinate traffic signals in both directions along an arterial without explicitly assigned offsets or constraints. The results also reveal C-MP's superiority to benchmark coordination strategies in reducing travel time, and fuel consumption both at the corridor level and the network level by balancing the negative impact imparted to vehicles in the minor direction.

*Keywords:* Max Pressure algorithm, adaptive traffic signal control, decentralized signal control, coordinated traffic signals

## 1. Introduction

Adaptive traffic signal controls (ATSC) have emerged as a promising solution to address urban traffic congestion. Centralized ATSC systems optimize traffic signal timings for a set of traffic signals simultaneously using a single central control unit (Gartner, 1983; Gettman et al., 2007; Head et al., 2006; Mirchandani and Head, 2001; P B Hunt et al., 1981). Unfortunately, these systems are generally not scalable due to computational complexity and data requirements involved. Decentralized ATSC systems are more computationally and data efficient as each intersection optimizes its signal plans without input or collaboration with others. One example that

*Corresponding author

*Email addresses:* tpa5285@psu.edu (Tanveer Ahmed), hfl5376@psu.edu (Hao Liu), gayah@engr.psu.edu (Vikash V. Gayah)

is growing in the research literature is the Max Pressure (MP) framework, which only requires local information on vehicle metrics and turning ratios at a given intersection to make signal timing decisions. Proposed initially for packet transmission in wireless systems (Tassiulas and Ephremides, 1990), MP applied to traffic signal control was first introduced as a decentralized ATSC by (Varaiya, 2013; Wongpiromsarn et al., 2012). The MP framework requires no knowledge of traffic demands and has been theoretically proven to be able to serve any demand at an intersection that can be feasibly served by any other signal control strategy. This latter property is known as *maximum stability*. Since its proposed application in traffic signal control, the MP control policy has been widely studied by researchers who proposed modifications to allow more flexible detection, controls and improved performances under different scenarios (Ahmed et al., 2024a; Anderson et al., 2018; Barman and Levin, 2022; Kouvelas et al., 2014; Levin et al., 2020; Lioris et al., 2016; Liu et al., 2024; Liu and Gayah, 2022, 2023, 2024; Mercader et al., 2020; Pumir et al., 2015; Ramadhan et al., 2020; Wang et al., 2022; Wei et al., 2019; Xu et al., 2022; Zoabi and Haddad, 2022a, 2022b).

One significant drawback of the MP framework is the lack of coordination between signal timings at adjacent intersections due to its decentralized nature. For traffic signals, coordination seeks to provide uninterrupted passage for a group of vehicles traveling together (i.e., a platoon) along a corridor with closely spaced intersections. The simplest coordination mechanism requires all signals to operate with the same cycle length and involves implementing an offset, which represents the time interval between the start of the coordinated phase at the upstream intersection and the start of the same phase at the downstream intersection. This is typically set to the free-flow-travel-time of vehicles on the link. Coordinating in this way significantly reduces the number of stops along the corridor ensuring uniform speeds and smooth flow (Berg et al., 1986). Fewer stops translate to reduced fuel consumption, lower pollutant emissions, and vehicle operating costs compared to stop-and-go traffic conditions (De Coensel et al., 2012). Studies have also shown that well-coordinated corridors also reduce the potential for vehicular conflicts, particularly rear end crashes, i.e., improve the overall safety performance (Guo et al., 2010; Lin and Huang, 2010; Yue et al., 2022; Zeng et al., 2015).

Numerous studies have proposed different coordination strategies for signalized intersections. MAXBAND and MAXBAND-86 maximize inbound and outbound bandwidth along an arterial through adjustments to cycle time, offsets, and phase sequences (Gartner et al., 1975; Little, 1966; Little et al., 1981; Morgan and Little, 1964). This concept was later extended to MULTIBAND (Gartner et al., 1991; Gartner et al., 1990) and MULTIBAND-96 (Stamatiadis and Gartner, 1997), which incorporate systematic traffic-dependent criteria into the optimization objective and can generate progression bands with varying widths using a mixed-integer linear programming (MILP) model. However, inadequate bandwidth to accommodate traffic demands can lead to the formation of residual queues, resulting in inefficient traffic flow. When these residual queues become substantial, an effective solution is to synchronize signal offsets with the progression of backward waves, as documented in (Chaudhary and Messer, 1993; Daganzo et al., 2018; Daganzo and Lehe, 2016; Pignataro et al., 1978; Quinn, 1992; Rathi, 1988; Sadek et al., 2022). This approach ensures that signal phases for a particular traffic movement are only initiated when the intersection is clear of any downstream queues that might obstruct it. Another branch of literature explored adaptive control methods – e.g., SCATS, RHODES, UTOPIA, and PAMSCOD – that can operate without fixed cycle lengths, allowing coordination to occur from shared optimization rules among neighboring intersections (Gartner, 1983; He et al., 2012; Mirchandani and Head, 2001; P B Hunt et al., 1981; P R Lowrie, 1982; Pavleski et al., 2017). However, most of these adaptive algorithms are centralized strategies that result in complex mixed integer linear programs (MILP) that are not scalable as the number of intersections to be coordinated grows. Self-organizing traffic lights (SOTL) are able to implicitly achieve a limited amount of coordination (Cesme and Furth, 2014; Lämmer and Helbing, 2008; Xie et al., 2011); however, unlike MP, SOTL does not consider downstream traffic properties and does not have theoretical guarantees of throughput. With the increasing popularity of artificial intelligence, methods such as machine learning, reinforcement learning and artificial neural networks have also been applied to coordinate signals (Abdoos, 2021; Chu et al., 2020; Park et al., 2001; Shoufeng et al., 2008; Wei et al., 2019; Zhang et al., 2019). Unfortunately, a limitation to applying these techniques is the learning process that takes many iterations of trial and error meaning its application in real-life is farfetched.

Thus, integrating signal coordination into the MP framework is of great research interest. A recent study (Xu, 2023) proposed "Smoothing-MP": an MP algorithm that has the ability to coordinate signals via a rule-based constraint. The proposed algorithm forces the downstream link in the coordinated direction to have a higher pressure when its upstream was just served and therefore increases the chance of the downstream link being served in the following time step. However, its performance is questionable when link lengths are asymmetrical or very long as platoons may not reach the downstream intersection within the following time step. Moreover, the proposed algorithm is unable to coordinate traffic in both travel directions simultaneously. Despite outperforming the original MP, its performance was also not compared against existing coordination algorithms.

In light of these gaps, this study proposes Coordinated Max Pressure (C-MP): a novel decentralized adaptive-coordinated traffic signal control using the MP framework. C-MP is built on the original acyclic MP algorithm that uses vehicle queues to identify the demand and supply on upstream and downstream links. The contribution is the integration of instantaneous space mean speeds (SMS) of the vehicles on upstream and downstream links to identify what portion of the vehicles are stopped or traveling in a freely flowing platoon. Specifically, C-MP provides a higher weight to larger upstream platoons to prioritize the movement and lower weights to platoons on downstream links that are likely to not disrupt available supply. By integrating this information, coordination is naturally provided in both travel directions along the arterial within the traditional decentralized MP framework. The study analytically proves that C-MP maintains the maximum stability property with no reduction in the stable region; i.e., the set of demands that can be served is not changed. This is also demonstrated via a stability analysis using micro-simulation, which shows that the C-MP can serve larger demands than several benchmark MP control polices, including the original MP (Q-MP), travel-time based MP (TT-MP), position-weighted back pressure (PWBP) and the rule-based MP proposed in (Xu, 2023) (Smoothing-MP). The simulation results also show that C-MP ensures coordination in both directions along a corridor without the need for explicitly assigning offsets. Compared to benchmark control policies, C-MP achieves lower travel time, results in fewer stops along a corridor and lower fuel consumption.

The remainder of this paper is as follows. Section 2 introduces the proposed C-MP control policy and provides the theoretical proof of maximum stability. Section 3 describes the simulation setup and the benchmark methods used to evaluate the performance of C-MP. Section 4 provides the results of the experiments are presented, including a comparison between the proposed method and the benchmark approaches. Finally, Section 5 highlights the findings and suggests directions for future work.

## 2. Method

This section introduces the control mechanism of the original MP, the proposed C-MP and its analytical properties.

### 2. 1.  Control mechanism of MP

The network model considered here consists of links and nodes. Each link denotes a unidirectional stretch of road connecting two nodes (i.e., intersections). At any given node, the upstream and the downstream links accommodate the flow of traffic into and out of the intersection, respectively. Figure 1 shows node $i$ along with its upstream link ($l$) and downstream link ($m$) in the eastbound direction. Any movement is defined by the pair of upstream and downstream links that allow vehicle transitions at an intersection; e.g., $(l, m)$ represents the eastbound through movement from link $l$ to link $m$ in Figure 1. The set of all upstream links at a node $i$ is denoted by $U(i)$, and $D(m) = \{n, o, p\}$ denotes the set of downstream links emanating from link $m$. The turning ratio, which is the fraction of traffic turning from link $l$ onto link $m$, is defined as $r(l, m)$. The maximum discharge rate of vehicles from an upstream link $l$ to a downstream link $m$ is denoted by the saturation flow, $c(l, m)$. The set of signal phases allowed by the signalized intersection is indicated by $\Phi_i$, where each individual phase $\phi$ allows a specific subset of movements, denoted by $L_i^{\phi}$.

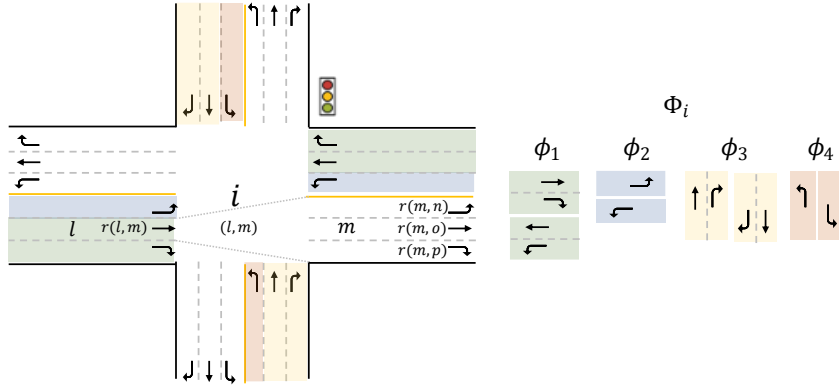

**Figure 1. Movements and turning ratios at an intersection**

The original MP proposed by (Varaiya, 2013)—referred to here as Q-MP—is an acyclic MP algorithm that measures vehicle queues and updates the signals at the end of discrete update intervals. The Q-MP algorithm follows three steps:

1. First, weights ($w$) are assigned to a movement by calculating the difference between a specific vehicular metric (e.g., vehicle queues, weighted vehicle queues, travel time, delay, etc.) associated with that movement and the average metric for its downstream movements:

$$w_q(l, m)(t) = w_q^{up} - w_q^{down}$$
$$= x(l, m)(t) - \sum_{n \in D(m)} x(m, n)(t) \times r(m, n)(t), \tag{1}$$

where $w_q(l, m)(t)$ denotes the weight of movement $(l, m)$ at time $t$ using the Q-MP; $w_q^{up}$ is the upstream metric of the weight; $w_q^{down}$ is the downstream metric of the weight; and $x(l, m)(t)$ denotes the number of queued vehicles on $(l, m)$ at time $t$. The upstream metric reflects the level of demand on a link, while the downstream indicates the amount of supply (or space) available to accommodate the upstream demand.

2. Second, the pressure ($P$) of a phase is calculated by aggregating the product of the weight and the saturation flow rate across all movements accommodated by that phase:

$$P_q^\phi = \sum_{(l,m) \in L_i^\phi} w_q(l, m) \times c(l, m), \ \forall \phi \in \Phi_i, \tag{2}$$

where $P_q^\phi$ denotes the pressure of phase $\phi$ using the Q-MP. This allows the algorithm to rank each phase served by the signalized intersection.

3. Finally, the phase with max pressure is activated ($S$):

$$S_q = \arg\max_{\phi \in \Phi_i} P_q^\phi \tag{3}$$

where $S_q$ serves as an indicator variable for the signal status on a given phase using the Q-MP. The acyclic structure of MP algorithms selects the phase with the maximum pressure to receive green for some discrete time interval, $\Delta T$.

## 2. 2. *Proposed C-MP algorithm*

A drawback of Q-MP is that it relies only on vehicle counts and fails to consider if and how vehicles are moving, which could lead to both platoons being broken or signal timing changes that disrupt progression. As an example, Figure 2 shows a platoon of vehicles traveling eastbound across a series of intersections on a major arterial with queued vehicles on the cross-streets in the minor direction. The illustration shows a snapshot $\Delta T$ time after the upstream movement at intersection $i$ is served and a signal update decision for the next interval is being made where the first three vehicles in the platoon exit link $l$ and join link $m$ to travel toward intersection $j$. At this instant, the Q-MP would record major movement as having a weight of one vehicle, since there are four vehicles upstream and three downstream. The minor approach at intersection $i$ has a higher weight (three vehicles) and thus would be called. Doing so breaks the platoon of seven vehicles (four on link $l$, three on link $m$), which would hinder coordination. The primary reason for this is that the Q-MP treats stopped and moving vehicles the same. So, for example, the three vehicles traveling on downstream link $m$ are seen as an obstacle to vehicles moving on the upstream link $l$ in the major direction. Further, arriving platoons maintain a consistent spacing near the critical density, while stopped vehicles are densely packed at jam density, resulting in fewer moving vehicles captured within the same detection length (see intersection $j$). This increases the likelihood of activating minor movements; however, it fails to recognize that calling the minor movement would cause the vehicles traveling at free flow to have to stop. This renders the Q-MP ineffective at detecting platoons and providing coordination along an arterial.

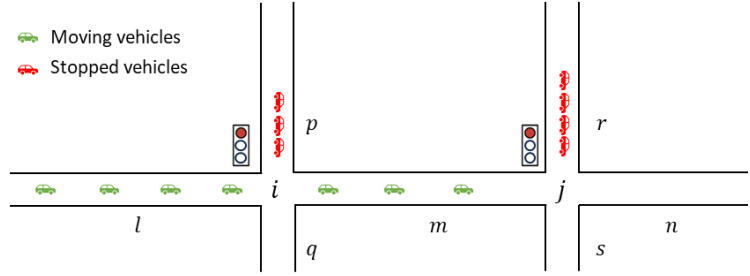

**Figure 2. MP's inability to detect moving vehicles upstream and downstream leads to disruptions to flow of platoons**

This problem can be addressed by capturing information on the condition of the vehicles near the intersection (specifically, whether they are moving or not moving) in the pressure metric. The rest of this section introduces the proposed Coordinated Max Pressure (C-MP) control policy that incorporates instantaneous space mean speed (SMS) into the weight calculation to recognize traffic conditions on links both upstream and downstream of an intersection. Doing so allows the C-MP to explicitly consider the arrival of a platoon on an upstream link or departure of a platoon on a downstream link, facilitating coordination even within the decentralized environment. Modifications to the upstream and downstream portions of the metric calculation in (1) are first described individually, then combined into the final C-MP algorithm.

### 2. 2. 1. Modified upstream metric

Proper detection of platoons upstream is a crucial element in ensuring the smooth progression of traffic along a coordinated corridor. The C-MP algorithm modifies the metric associated with the upstream movement $(l, m)$ to detect and prioritize the movement of platoons as follows:

$$w_c^{up} = x(l,m)(t) \times \left(1 + \beta \frac{\bar{v}_{(l,m)(t)}}{v_f}\right) = w_q^{up} \times \left(1 + \beta \frac{\bar{v}_{(l,m)(t)}}{v_f}\right),\qquad(4)$$

where $w_c^{up}$ is the upstream metric associated with the weight of movement $(l, m)$ using the C-MP; $\bar{v}_{l,m(t)}$ is the SMS of vehicles on movement $(l, m)$ at time $t$; $v_f$ is the free-flow-speed; and, $\beta$ is a tuning factor that takes values between $[0, \frac{k_j}{k_c} - 1]$.

The modification essentially increases the original upstream weight by a factor of $\left(1 + \beta \frac{\bar{v}_{(l,m)(t)}}{v_f}\right)$. This factor increases with the measured SMS of vehicles on the upstream link, which is more indicative of vehicles traveling in a platoon as opposed to vehicles queued at the intersection. This modification ensures that the algorithm prioritizes platoons, while also serving longer queues on competing phases as they build up, and $\beta$ can be used to tune how much platoons are prioritized. When $\beta = 0$, the system reverts to the Q-MP where the average speed does not influence the weight of the movement. As $\beta$ increases, the system weighs more heavily the movements with faster moving upstream vehicles; i.e., those in a platoon. The upper bound of $\beta$ ensures that the adjusted values do not unrealistically exceed what would be possible under jam conditions. This means that the scaling of the number of free-flowing vehicles at the critical density does not exceed the equivalent number of stopped vehicles that would occupy the same space at jam density. Note that if all vehicles on the movement are stopped, $\bar{v}_{(l,m)(t)} = 0$ and the expression also reverts back to that in the Q-MP.

### 2. 2. 2. Modified downstream metric

The presence of stopped vehicles downstream hinders the progression of platoons through a corridor of signalized intersections. However, if the downstream vehicles are moving, additional space may be available to accommodate upstream vehicles without them needing to stop. Therefore, to prevent abrupt signal changes and allow a platoon to continue progressing, the weight of downstream vehicles traveling in free-flow states can be reduced. To account for this traffic condition, the downstream metric in the proposed C-MP is:

$$w_c^{down} = \sum_{n \in D(m)} x(m,n)(t) \times r(m,n)(t) \times \left(1 - \alpha \frac{\bar{v}_{(m,n)(t)}}{v_f}\right) = w_q^{down} \times \left(1 - \alpha \frac{\bar{v}_{(m,n)(t)}}{v_f}\right),$$

(5)

where $w_c^{down}$ is the downstream metric associated with the weight of the C-MP; and $\alpha$ is a tuning factor that takes values between [0,1].

The modification decreases the impact of the original downstream weight by a factor $\left(1 - \alpha \frac{\bar{v}_{m,n(t)}}{v_f}\right)$. This factor decreases with the measured SMS of vehicles on the downstream link, which is indicative of vehicles traveling in a platoon (as opposed to vehicles queued and taking space downstream). The modification ensures that platoons of vehicles downstream that are moving count less in the weight calculation than queued/stopped vehicles, and $\alpha$ can be used to tune the amount of the reduction. If $\alpha = 0$, this term becomes 1, meaning that the current speed of traffic does not affect the downstream metric. All vehicles, regardless of their speed, are fully accounted for in the metric and the metric reverts to the original Q-MP. As $\alpha$ increases toward 1, the impact of vehicles traveling at $v_f$ is gradually decreased in the downstream metric by $(1 - \alpha)$. Thus, at $\alpha = 1$, the algorithm would only consider the number of stopped vehicles and all vehicles moving at $v_f$ would be disregarded. Like the upstream modification, when all vehicles are stopped, the expression reduces to the downstream metric of Q-MP and all downstream vehicles are considered.

### 2. 2. 3. Control mechanism of C-MP

Combining the proposed modification to the upstream (4) and downstream (5) to (1), the C-MP control policy calculates the weight of a movement $(l,m)$ at the end of each discrete update interval at time $t$ using (6).

$$w_c(l,m)(t) = x(l,m)(t) \times \left(1 + \beta \frac{\bar{v}_{(l,m)(t)}}{v_f}\right) - \sum_{n \in D(m)} x(m,n)(t) \times r(m,n)(t) \times (1 -$$
$$\alpha \frac{\bar{v}_{(m,n)(t)}}{v_f})$$

(6)

Together, the proposed tuning parameters $\alpha$ and $\beta$ can be used to control the priority to platoons and the strength of coordination imposed along the arterial. Increasing both would provide more priority to movements serving platoons along the arterial that coordination is desired. The pressure of a phase $\phi$ at node $i$ using C-MP is calculated similar to Q-MP, except using the weight calculated from (6):

$$P_c^{\phi} = \sum_{(l,m) \in L_i^{\phi}} w_c(l,m) \times c(l,m), \ \forall \phi \in \Phi_i. \tag{7}$$

Finally, C-MP selects the phase $\phi$ with the max pressure considering the number of vehicles and their average speeds:

$$S^* = \arg\max_{\phi \in \Phi_i} P^{\phi}. \tag{8}$$

The next section proves that the C-MP maintains the maximum stability property that is desirable in the MP algorithm.

### 2. 3. Maximum stability of C-MP

A signal control policy is stable if the number of vehicles in the network are upper bounded, i.e., they do not keep growing over time. Maximum stability refers to the property that the policy can serve a traffic demand if this demand can be accommodated by any admissible control strategy. In order to theoretically prove the control strategy is stable, similar assumptions are made and steps are followed to those in (Liu and Gayah, 2022; Varaiya, 2013). This includes the adoption of the store-and-forward model for the evolution of vehicles on a link which assumes a point queue model (i.e., the spatial extent of a queue is ignored) and that queue capacities are infinite. The following sub-section includes the assumptions, propositions, and definitions pertaining to the proof. Note that these assumptions are only made for the proof and do not impact the application of the model as proposed in (6-8).

**Assumption 1.** *A vehicle may only be in either a state of free-flow or jam when traveling in a network of signalized intersections.*

According to Assumption 1, vehicles are in a state of jam when they stop at a red light. All other vehicles that arrive from an upstream source travel at $v_f$, and these vehicles may arrive randomly or together in a platoon. At any time, vehicles on a link may exist in one of these two states: stopped or moving as shown in Figure 3. The total number of vehicles on a movement $(l,m)$ at time $t$ can then be described as the sum of stopped and moving vehicles as follows:

$$x(l,m)(t) = x_s(l,m)(t) + x_m(l,m)(t), \tag{9}$$

where $x_s(l,m)(t)$ and $x_m(l,m)(t)$ are the number of stopped and moving vehicles on movement $(l,m)$ at time $t$, respectively. The number of stopped vehicles increases if moving vehicles join the back of the existing queue and decreases as the queue is served and vehicles depart the link. On the contrary, the number of moving vehicles decreases as they join the back of an existing queue on that link or if they leave the link while traveling at $v_f$. The number of moving vehicles increases on an entry link when there is exogenous demand on that link and on an internal link when the outflow from an upstream link joins it. The traffic evolution of stopped and moving vehicles is described in Section 4.1. of (Liu and Gayah, 2022) and not repeated here to avoid repetition. In addition, according to Proposition 2 of (Liu and Gayah, 2022), the number of moving vehicles on a link is upper bounded by a constant. This is critical for establishing the theoretical proof of maximum stability of the proposed C-MP.

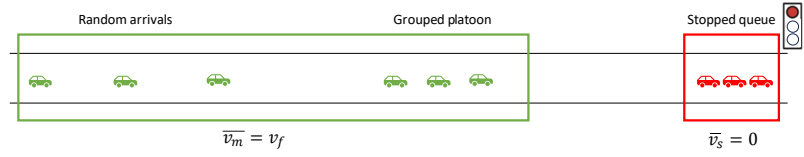

**Figure 3. States of vehicles on a link**

**Proposition 1.** *The ratio of instantaneous space-mean-speed to the free-flow-speed of vehicles on a link is equal to the proportion of moving vehicles on the link.*

**Proof.** Based on Assumption 1, at any time $t$, the speed of all moving vehicles $x_m(l,m)(t)$ is $v_f$, and the speed of all stopped vehicles $x_s(l,m)(t)$ is 0. The SMS of all vehicles on movement $(l,m)$ at time $t$, $\bar{v}(l,m)(t)$

$$= \frac{x_m(l,m)(t) \times v_f + x_s(l,m)(t) \times 0}{x_m(l,m)(t) + x_s(l,m)(t)}$$

$$= \frac{x_m(l,m)(t) \times v_f}{x(l,m)(t)} . \tag{10}$$

Therefore, the ratio of the SMS to the FFS of vehicles on a link at any time can be written as,

$$\frac{\bar{v}(l,m)(t)}{v_f} = \frac{x_m(l,m)(t)}{x(l,m)(t)} , \tag{11}$$

which is equal to the proportion of moving vehicles to the total number of vehicles on the link.

Using (11) from Proposition 1, (6) can be decomposed as follows:

$w_c(l,m)(t)$

$$= x(l,m)(t) \times \left(1 + \beta \times \frac{x_m(l,m)(t)}{x(l,m)(t)}\right) - \sum_{n \in D(m)} x(m,n)(t) \times r(m,n)(t) \times (1 - \alpha \times \frac{x_m(m,n)(t)}{x(m,n)(t)})$$

$$= \left[x(l,m)(t) - \sum_{n \in D(m)} x(m,n)(t) \times r(m,n)(t)\right] + \left[x(l,m)(t) \times \beta \times \frac{x_m(l,m)(t)}{x(l,m)(t)} + \sum_{n \in D(m)} x(m,n)(t) \times r(m,n)(t) \times \alpha \times \frac{x_m(m,n)(t)}{x(m,n)(t)}\right]$$

$$= \left[x(l,m)(t) - \sum_{n \in D(m)} x(m,n)(t) \times r(m,n)(t)\right] + \left[\beta x_m(l,m)(t) + \alpha \sum_{n \in D(m)} x_m(m,n)(t) \times r(m,n)(t)\right] \tag{12}$$

From (1), the term in the first square brackets can be rewritten as the weight of Q-MP:

$$w_c(l,m)(t) = w_q(l,m)(t) + \left[\beta x_m(l,m)(t) + \alpha \sum_{n \in D(m)} x_m(m,n)(t) \times r(m,n)(t)\right] \tag{13}$$

Therefore, it is possible to alternatively calculate the weight of a movement for C-MP as the sum of the weights calculated for Q-MP and the proportion of moving vehicles upstream and downstream multiplied their respective tuning parameters.

**Definition 1.** *A demand $d$ is feasible if there exists a signal control sequence $S(t)$ such that:*

$$\overline{d}(l,m) \leq \overline{S}(l,m)c(l,m) \quad \forall(l,m), \tag{14}$$

where $\overline{d}(l,m)$ is the average external demand of movement $(l,m)$, $\overline{S}(l,m)$ is the average proportion of update intervals that movement $(l,m)$ is activated, and $c(l,m)$ is the average saturation flow for movement $(l,m)$.

The set of demands satisfying (14), denoted by $\mathcal{D}$, is called feasible demand region, and $\mathcal{D}^0$ is used to indicate the interior of $\mathcal{D}$. Therefore, a demand scenario is feasible if there exists a signal control sequence from which the average service rate for all movements in the long run is higher than the average arrival rate (Varaiya, 2013).

**Definition 2.** *A control sequence $S(t)$ is stable in the mean if the average number of vehicles in the network, $\frac{1}{T}\sum_{t=1}^{T}\sum_{l,m}\mathbb{E}[x(l,m)(t)]$, is finite for all $T$.*

It has been shown that stable control sequences exist if and only if the demand is feasible. The proof can be found in (Varaiya, 2013).

**Theorem 1.** *The C-MP is stable if $d \in \mathcal{D}^0$.*

**Proof:** According to (Varaiya, 2013), a signal control policy can be proved to be stable if there exists $k < \infty$ and $\epsilon > 0$ such that, following inequality holds under the C-MP control policy:

$$E\{|X(t+1)|^2 - |X(t)|^2|X(t)\} \leq k - \epsilon|X(t)|, t = 1,2,\dots, \tag{15}$$

where $|X(t)|^2$ is the vector containing the sum of squares of all queue lengths, i.e., $|X(t)|^2 = \sum_{l,m}\big(x(l,m)\big)^2$.

Taking the unconditional expectation and summing over $t = 1,2,\dots T$ yields:

$$E|X(t+1)|^2 - E|X(1)|^2 \leq kT - \epsilon\sum_{t=1}^{T}E|X(t)|,$$

$$\therefore \epsilon\frac{1}{T}\sum_{t}^{T}E|X(t)| \leq k + \frac{1}{T}E|X(1)|^2 \leq k + \frac{1}{T}E|X(1)|^2, \tag{16}$$

which would denote that the average number of vehicles in the network is upper bounded. In order to prove (15), let the change in the number of vehicles in the network in consecutive signal update intervals between $t$ and $t+1$ be denoted by vector $\delta$, where $\delta = X(t+1) - X(t)$. Therefore:

$$|X(t+1)|^2 - |X(t)|^2 = 2X(t)^T\delta + |\delta| = 2\theta + \lambda \tag{17}$$

Thus, it is required to prove that both $\theta$ and $\lambda$ are upper bounded.

**Lemma 1.** *$\theta$ is upper bounded.*

Following the routing and flow conservation principles and the steps from (A.3)-(A.4) in (Varaiya, 2013)

$$E\{\alpha|X(t)\} = \sum_{l\in\mathcal{L}}[d_l r(l,m) - E\{\min\{C(l,m)(t)S^*(l,m)(t),x(l,m)(t)\}|X(t)\}]w_q(l,m)(t)$$

$$= \theta_1 + \theta_2, \tag{18}$$

Where:

$$\theta_1 = \sum_l [d_l r(l,m) - c(l,m)S^*(l,m)(t)]w_q(l,m)(t) \text{ and} \tag{19}$$

$$\theta_2 = \sum_{l \in \mathcal{L}} [c(l,m) - E\{\min\{C(l,m)(t), x(l,m)(t)\}|X(t)\}]S^*(l,m)(t)w_q(l,m)(t). \tag{20}$$

$w_q(l,m)(t)$ is the weight of movement $(l,m)$ calculated using the original MP (Q-MP) defined in (1). However, $S^*(l,m)(t)$ is the signal state of the phase serving movement $(l,m)$ at time $t$ according to the weight $w_c(l,m)(t)$ calculated using C-MP control policy defined in (13).

**Lemma 1.1.** *$\theta_2$ is upper bounded.*

From (20), it is evident that:

$$c(l,m) - E\{\min\{C(l,m)(t), x(l,m)(t)\}|X(t)\}$$
$$= \begin{cases} 0 & \text{if } x(l,m)(t) \geq C(l,m)(t+1) \\ \leq c(l,m)(t) & \text{otherwise} \end{cases}$$

Furthermore, $S^*(l,m)(t)$ is a binary function with values $[0, 1]$ and, from (1), $w_q(l,m)(t) \leq x(l,m)(t)$. Thus, $\theta_2 \leq c(l,m)\bar{C}(l,m)$; i.e., it is upper bounded by a constant where $c(l,m)$ and $\bar{C}(l,m)$ are the mean and upper bound of the saturation flow, respectively.

**Lemma 1.2.** *$\theta_1$ is upper bounded.*

From (19):

$$\theta_1 = \sum_l [d_l r(l,m) - c(l,m)(t)S^*(l,m)(t)]w_q(l,m)(t)$$

$$= \sum_l [d_l r(l,m) - c(l,m)(t)S^*(l,m)(t)][w_c(l,m)(t) - [\beta x_m(l,m)(t) + \alpha \sum_{n \in D(m)} x_m(m,n)(t) \times r(m,n)(t)]]$$

$$= \sum_l [d_l r(l,m) - c(l,m)(t)S^*(l,m)(t)]w_c(l,m)(t)$$

$$- \sum_l [d_l r(l,m) - c(l,m)(t)S^*(l,m)(t)][\beta x_m(l,m)(t) + \alpha \sum_{n \in D(m)} x_m(m,n)(t) \times r(m,n)(t)]$$

$$= \theta_{11} + \theta_{12} \tag{21}$$

**Lemma 1.2.1** *$\theta_{12}$ is upper bounded.*

Since it was previously proven from Proposition 2 in (Liu and Gayah, 2022) that the number of moving vehicles on a link, $x_m(l,m)(t)$ is upper bounded, and both $\alpha$ and $\beta$ are non-negative and finite, it is evident that $\theta_{12}$ is also bounded.

**Lemma 1.2.2** $\theta_{11}$ *is upper bounded.*

Since, $d \in D$ , there exists a signal control matrix $\Sigma^+ \in co(S)$ and $\epsilon > 0$ such that, $c(l,m)\Sigma^+(l,m) > d_l r(l,m) + \epsilon \ \forall(l,m)$. Here $co(S)$ is used to denote the convex hull of all possible signal timings. Therefore, for any fixed $t$, there also exists $\Sigma \in co(S)$ such that $0 \leq \Sigma \leq \Sigma^+$ and:

$$c(l,m)\Sigma(l,m)(t) = \begin{cases} d_l r(l,m)(t) + \epsilon, & if \ w_c(l,m)(t) > 0 \\ 0, & otherwise \end{cases}. \tag{22}$$

Since C-MP selects the phase with the maximum pressure defined in (7-8), $S^*$ maximizes the term of $c(l,m)(t)S^*(l,m)(t)w_c(l,m)(t)$:

$$\theta_{11} \leq \sum_l[d_l r(l,m) - c(l,m)(t)\Sigma(l,m)(t)]w_c(l,m)(t)$$

$$\leq -\epsilon \sum_l w_c^+(l,m)(t) + \sum_l[d_l r(l,m)]w_c^-(l,m)(t)$$

$$\leq -\epsilon|w_c(l,m)(t)|, \tag{23}$$

where $w_c^+ = \max\{w_c, 0\}$ and $w_c^- = \max\{-w_c, 0\}$.

From (13), it can be seen that, $w_c(l,m)$ is a linear combination of $w_q(l,m)$ and $X_m = \{x_m(l,m)\}$. (Varaiya, 2013) proved that based on the 1:1 properties of the function and the routing probabilities $\{r(l,m)\}$ there exists a constant $\eta > 0$ such that, $\sum_l|w_q(l,m)| \geq \eta|X(t)|$ . In addition, since the number of moving vehicles is upper bounded (Liu and Gayah, 2022) while both $\alpha$ and $\beta$ are non-negative and finite,

$$\sum_l|w_c(l,m)| \geq \sum_l|w_q(l,m)| \geq \eta|X(t)| \tag{24}$$

Therefore, combining (23) and (24) yields $\theta_{11} \leq -\epsilon\eta|X(t)|$. Hence, lemma 1.2.2 is proved, i.e., $\theta_{11}$ is upper bounded.

**Lemma 2.** $\lambda$ *is upper bounded.*

Based on the evolution of vehicle queues defined by the store-and-forward model, the difference in the number of vehicles in the network between two consecutive time steps is upper bounded by the maximum value of the demand in the network. Therefore, $\lambda = |\delta|^2$ is upper bounded a constant. This has been proven in (Varaiya, 2013) and not repeated here.

Thus, the upper bound on both $\theta$ and $\lambda$ is established, and Theorem 1 is proved.

## 3. Simulation Setup

The performance of the proposed C-MP algorithm is tested within a simulation environment. This section describes the simulation details, as well as other algorithms used as benchmarks to compare its performance.

### 3. 1. Network setup

The AIMSUN micro-simulation software was used for the simulation tests due to its ability to accurately model traffic dynamics in a network (Barceló and Casas, 2005) and ease of programming signal control algorithms including the MP (Ahmed et al., 2024b). Since the objective was to demonstrate the proposed C-MP can provide coordination along an arterial, simulation tests were carried out on an arterial network consisting of 1 major corridor in the east-west direction, 8 minor links in the north-south direction and a series of 8 signalized intersections; see Figure 4. Internal links were asymmetrical with varied lengths between 150 m and 300 m, while the speed limit was set to 50 km/hr. As shown in Figure 5, each major link (E-W direction) had three lanes on each approach to accommodate dedicated left, through and right turning movements, while minor links (N-S direction) had two lanes per approach: one shared through and right turn lane and one dedicated left turn lane. Every signalized intersection had four potential phases: through and right share a phase while left turns had their own dedicated phases on each of the N-S and E-W directions; see Figure 6.

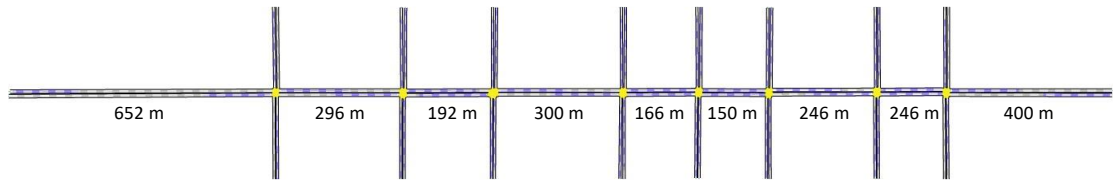

**Figure 4. Simulated arterial network**

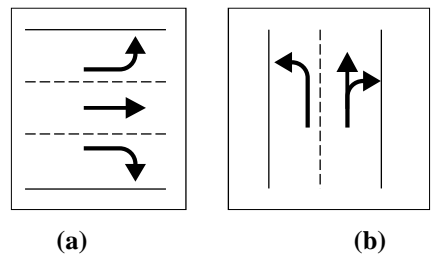

**(a)**                **(b)**

**Figure 5. Lane configuration (a) E-W direction; (b) N-S direction**

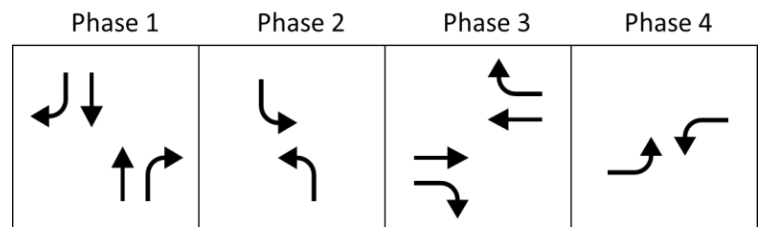

**Figure 6. Phase configuration**

## 3. 2. Scenario setup

Origins and destinations were placed at all entry/exit links. An OD matrix was constructed such that the demand on each of the entry links in the major direction was 5 times higher than the demand on each of the entry links in the minor direction. Since signal coordination is effective only when there is substantial traffic that travel end-to-end along a coordinated corridor, 60% of the vehicles entering the network through the major direction were assumed to travel end-to-end. Two distinct scenarios were simulated characterized by the level of demand: a high-demand scenario with a total input flow of 7,656 vehicles/hour and a medium demand scenario with a total input flow of 6,336 vehicles/hour.

Due to the asymmetric nature of the network, the duration of the signal update interval, $\Delta T$ was set to six seconds based on the shortest link in the arterial to ensure that platoons would be detected sufficiently upstream allowing the signals to change in time to allow them to progress through the intersection without stopping. To understand how tuning parameters $\alpha$ and $\beta$ impact the performance of C-MP, a range of different scenarios were simulated for both demand cases. Specifically, $\alpha$ values from 0 to 1 and $\beta$ values from 0 to 4 in increments of 0.1 were tested, where $\alpha = \beta = 0$ represents the Q-MP. The simulated links in the network have a ratio of $k_j/k_c = 5$; hence, the upper bound on $\beta$ was set to 4. Each configuration was simulated with 10 random seeds to ensure robust and comprehensive results.

### 3. 3.  Benchmark control methods

The performance of the C-MP algorithm was compared to five well established control methods:

- Original MP (Q-MP)
- Travel Time MP (TT-MP)
- Position Weighted Back Pressure (PWBP)
- Rule-based coordinated MP (Smoothing-MP)
- Sydney Coordinated Adaptive Traffic System, (SCATS-L)

The original MP (Q-MP) was used as the first benchmark. To compare the performance against an MP variant that uses a time averaged metric over the duration of the update interval, the travel-time based MP (TT-MP) inspired from (Mercader et al., 2020) was also selected as a benchmark. TT-MP utilizes the travel time of vehicles on upstream and downstream links, measured over the set update interval, to calculate the weight of a movement. The third benchmark method is the Position-Weighted-Back-Pressure (PWBP) proposed by (Li and Jabari, 2019). This is a variant of the MP that relies on instantaneous vehicle queues, but considers the spatial distribution of vehicles along the road to further weigh the upstream and downstream metrics. PWBP assigns weights to movements based on the sum of normalized distances of all vehicles, with distances calculated relative to the link length. Therefore, vehicles closer to the intersection are given a higher weight.

The other MP based benchmark method—referred to as "Smoothing-MP" in (Xu, 2023)— is a rule-based MP that incorporates traffic signal coordination. Whenever an upstream phase in the coordinated direction is served, this algorithm adds a "smoothing weight" to the pressure of the coordinated phase downstream to increase the chances of serving the phase. The value of the smoothing weight can control how much pressure is manually added to the downstream.[1] To replicate the best configuration of the Smoothing-MP in (Xu, 2023), a smoothing weight of 20,000 was used to increase the pressure of subsequent downstream phases in the eastbound direction.

---

[1] Note that this smoothing weight can only be applied to one direction along the arterial and thus would only provide coordination in one direction. Applying to both directions simultaneously to coordinate bi-directional traffic would result in an infinite activation of the coordinated phase on all intersections in the corridor.

The final benchmark method was a variant of the widely used Sydney Coordinated Adaptive Traffic System, (SCATS-L) from (Zhang et al., 2013). Unlike the acyclic MP algorithms, this is a cyclic adaptive-coordinated traffic signal control algorithm that optimizes cycle length and split time based on the measured degree of saturation (DoS) of each phase. The DoS at the first intersection of a coordinated corridor (termed as the master node) is calculated as the ratio of the traffic volume on the incoming links and the split time of a phase which is then used to set the cycle time at all the other intersections in the system. The split time of each phase is determined by proportionally dividing the cycle length by the degree of saturation of each phase. SCATS-L is well known to coordinate consecutive traffic signals by setting fixed offsets and maintaining the same cycle length, thus creating "green waves" for smoother traffic progression. The westernmost intersection was set as the master node and an offset was implemented on adjacent intersections in the eastbound direction equal to the FFTT of the link. Therefore, this control plan was designed to coordinate traffic signals in only the eastbound direction.

### 3. 4. Evaluation metrics

Several performance measures were used to evaluate the performance of each signal control strategy, including:

- Travel time
- Fuel consumption
- Network accumulation to assess stability

These were computed both globally (i.e., on the whole network) and along the arterial corridor (including only vehicles traveling end to end along major corridor). Note that fuel consumption depends on the state of vehicles at each instant that includes idling, cruising, accelerating and decelerating. While frequent stops require vehicles to accelerate to the free flow speed causing more fuel consumption, fewer stops allow a vehicle to cruise at its desired speed which leads to improved fuel efficiency. AIMSUN determines the state of vehicles at each simulation time step and uses the formula from (UK DoT, 1994) to calculate the fuel consumption. The value of the constants used to calculate the fuel consumption of vehicles is summarized in Table 1.

**Table 1. Fuel consumption model**

| State | Value |
|---|---|
| Idling / Decelerating | $0.333\ ml/s$ |
| Accelerating | $C1 = 0.260$ <br> $C2 = 0.420$ |
| Cruising <br> (Akcelic, 1982) | $V_m = 50\text{km/hr}$ <br> $K1 = 0.0991$ <br> $K2 = 0.008743$ |

In addition, trajectories of vehicles traveling end-to-end in the major direction were visualized using time-space diagrams. These diagrams are helpful in identifying whether the control algorithms naturally lead to platoon and green wave formation throughout the simulated corridor. Finally, a simulation analysis of the stability of each of the control methods was carried out by examining whether the total number of vehicles in the network remained stable or continued to grow with time under constant demands.

## 4. Results

This section provides the results of the microsimulation analysis and compares different performance measures of C-MP against the benchmark control methods.

### 4. 1. Recognition of traffic conditions

The C-MP algorithm introduces weighing factors to the upstream and downstream metrics that allow it to identify vehicle platoons and change signal timings to provide priority for these platoons along the corridor as opposed to Q-MP. To visually inspect this property, the network was first simulated with medium demand and fixed-time signal controls until the C-MP ($\alpha = 1, \beta = 1$) and Q-MP algorithms were activated after time = 430 seconds. Figure 7 illustrates a portion of the time-space diagrams for vehicles traveling through the corridor in the eastbound direction between 100-800 seconds. Horizontal lines show the locations of three intersections and the signal status of the phase serving the through movement in the east-west direction. It is evident that C-MP facilitates the smooth flow of vehicles across the corridor (Figure 7a). When the C-MP is activated at time 430 seconds, the upstream vehicles are free-flowing and part of a platoon; thus, C-MP increases their impact on the weight calculation and continues to serve the movement until the platoon dissipates. The vehicles downstream are also traveling at the free-flow-speed. As a result, the downstream factor discounts their presence in the weight calculation causing minimal reduction to the weight of the upstream movement. Furthermore, most residual queues from the fixed time control period were cleared before the upstream platoons arrived. Thus, vehicles were able to freely travel through the segment, as denoted by the constant slope of most trajectories.

On the contrary, under the Q-MP control policy (Figure 7b), vehicles encounter frequent stops due to repeated phase changes. Specifically, the presence of (moving) downstream vehicles results in a high downstream metric. In addition, moving vehicles upstream do not receive any priority; hence, the signal serves the competing phases with higher weights. This leads to a lower throughput on the corridor compared to C-MP.

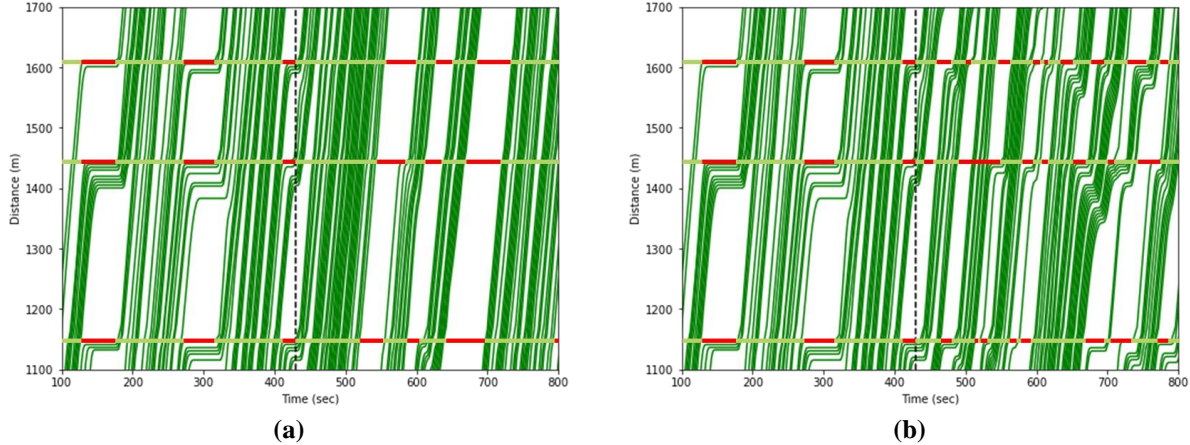

**Figure 7. Time-space diagram showing vehicle trajectories of (a) C-MP; (b) Q-MP**

### 4. 2.  Impact on travel time

Figure 8 shows the total network travel time when the C-MP strategy is implemented for the range of $\alpha$ and $\beta$ values tested. The shaded area around the curves represents the confidence interval associated with one standard error across the 10 random seeds that were simulated. The performance of the baseline methods are included as horizontal lines since their performance was not impacted by $\beta$: the solid line denotes Q-MP, the dotted line denotes the PWBP, the dashed line denotes the TT-MP, and the dashed-dotted line denotes the Smoothing-MP. SCATS-L is not shown since the travel time was much higher than the range provided in the figure; as it is a cyclic strategy, it was not as flexible and thus the travel time was the worst of the strategies tested.

The results reveal that the performance of the C-MP algorithm changes with respect to the tuning parameters depending on the demand levels. At medium demands, total travel time is fairly insensitive to $\alpha$ and increases with $\beta$ since increased priority to upstream platoons increases the delay for vehicles in the minor direction. At high demands, however, total travel time is insensitive to $\alpha$ only for larger $\beta$ values; for $\beta < 0.5$, the performance changes significantly with $\alpha$. At these lower $\beta$ values, C-MP relies primarily on the downstream weighing factor to implicitly provide coordination based on downstream traffic conditions. However, at higher $\beta$ values, platoons are explicitly detected and prioritized regardless of the traffic condition downstream, hence, increasing $\alpha$ results in insignificant change in travel time.

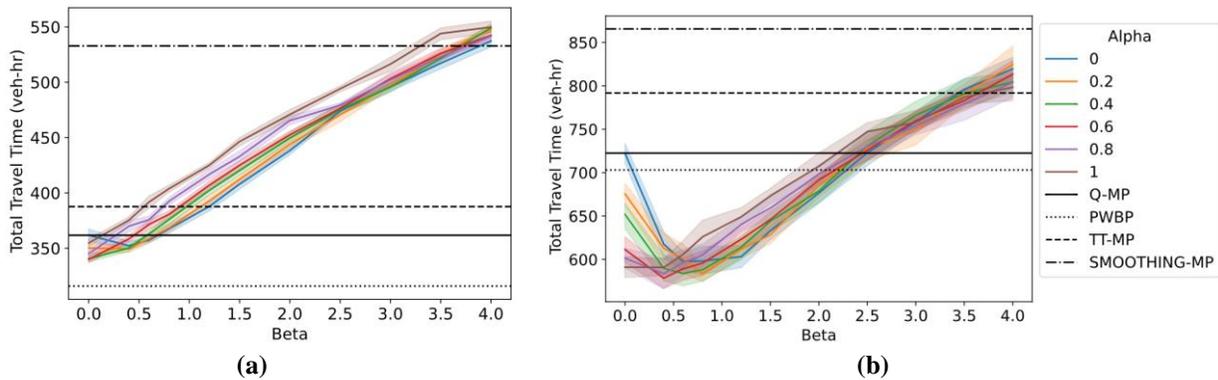

**Figure 8. Total network travel time: (a) medium demand; b) high demand**

Under higher demands, C-MP provides lower total network travel times than the comparison methods. For medium demands, the PWBP provides the lowest network travel times, and C-MP provides the second lowest travel times for certain combinations of $\alpha$ and $\beta$. However, C-MP better prioritizes arterial traffic compared to the PWBP; this is illustrated in Figure 9, which provides the travel time of vehicles on the arterial only. Further, C-MP and Smoothing-MP perform similarly at medium demands, and C-MP outperforms all benchmark control policies in terms of arterial travel time at high demands. Overall, improvements in corridor travel time is observed as $\beta$ increases to 1. The magnitude of this improvement is higher at the high demand level compared to the medium demand which explains the convex shape of Figure 8b where the total travel time first improves until beginning to rise. The TT-MP performs poorly in both demand scenarios due to the algorithm's reliance on activating the phase with the maximum vehicle travel time. This requires vehicles to experience delay before being receiving green, which is counterintuitive to promoting a continuous flow of traffic.

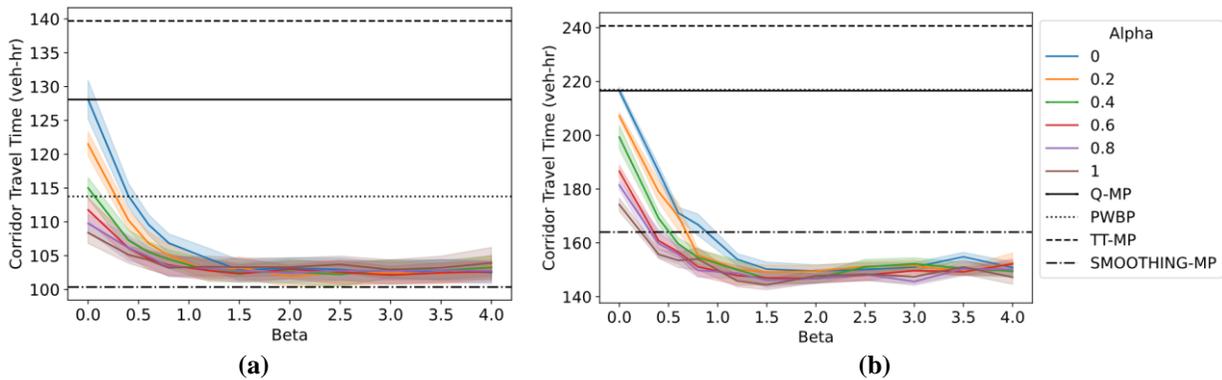

**Figure 9. Total travel time along corridor: (a) medium demand; b) high demand**

The travel time of vehicles along the corridor comprises vehicles traveling end-to-end in both the eastbound and westbound directions, which is illustrated separately in Figure 10 and Figure 11, respectively. Smoothing-MP outperforms C-MP in terms of travel time of vehicles in the coordinated eastbound direction for both demands levels as seen from Figure 10. This is a result of the smoothing weight that is only applied to increase the weight of downstream movements in the eastbound direction whenever the upstream movement has been served. By comparison, Figure 11 reveals that the Smoothing-MP significantly increases the travel time to vehicles travelling in the opposing westbound direction compared to C-MP, since coordination in this direction is not considered (and in fact hindered by coordination in the eastbound direction). However, C-MP is able to effectively coordinate signals in both directions along the corridor simultaneously.

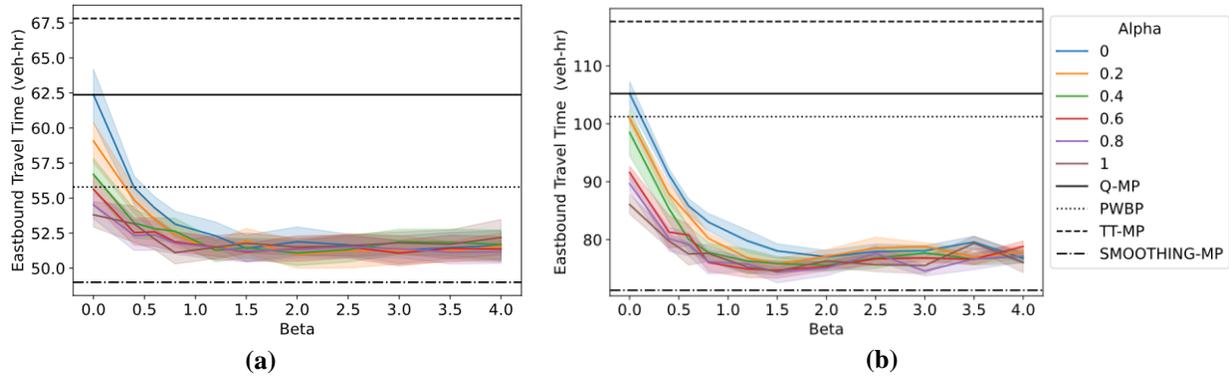

**Figure 10. Total travel time in eastbound direction: (a) medium demand; b) high demand**

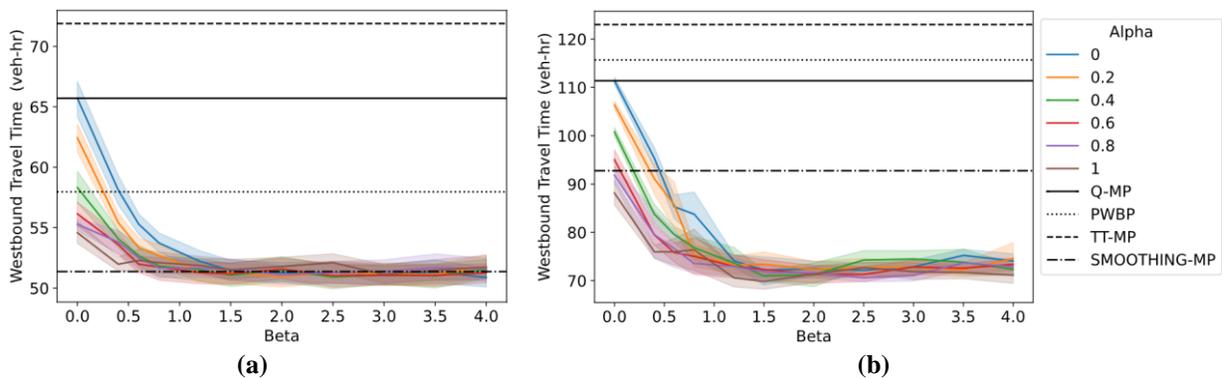

**Figure 11. Total travel time in westbound direction: (a) medium demand; b) high demand**

To further investigate the performance of each algorithm on the ability to provide coordination, vehicle trajectories were extracted from the medium demand case when each method was implemented; see Figure 12. The green lines plot the trajectories of individual vehicles that travel across the entire stretch of the corridor while gray horizontal dashed lines denote the locations of each signalized intersection. First, it is evident that under the C-MP control policy, distinct green-waves of vehicles are visible in both eastbound and the westbound directions (Figure 12a-b). This is evidence of well-coordinated signals that serve platoons in both directions until they fully cross the corridor. Since C-MP does not have external rules or fixed offsets to guarantee coordination, there are a few intersections where these platoons experience stops. However, the widths of the bands are consistent throughout the corridor, indicating that the platoons do not break apart. Platooning can be seen forming at the upstream-most intersection. Once released, this platoon is generally maintained throughout the corridor until the entire queue is fully served.

Figure 12b-h, show the trajectories for vehicles traveling eastbound and westbound under the Q-MP, TT-MP and PWBP respectively. None of these algorithms have a mechanism for coordinating traffic signals hence no progression is visible, and platoons released from an upstream intersection encounter stops at most downstream intersections. By comparison, Smoothing-MP is able to coordinate downstream traffic signals along the eastbound direction (Figure 12i) since the phase serving the eastbound-through movement is activated immediately after this phase has been activated on an upstream intersection. Therefore, downstream signals generally turn green before the head of a platoon arrives. However, some platoons are broken apart, especially on longer links. This is because the recommended update interval for Smoothing-MP is equal to the FFTT of a block length in a symmetric network, whereas the asymmetric structure of the simulated network means each link has a unique free-flow-travel-time. As a result, vehicles near the end of a platoon are unable to discharge and often form residual queues. On the contrary, vehicles traveling westbound only rely on the phase activation of the eastbound-through movement. Since their movements do not receive any explicit priority, progression is not visible in Figure 12j.

Finally, Figure 12k-l illustrate the time-space diagrams of vehicles using SCATS-L system. Although clear bands of green-waves corresponding to vehicles traveling eastbound (Figure 12k) are visible, the traffic signals on the westbound direction are not coordinated. Hence, platoons encounter frequent stops and wait until the next cycle to be served (Figure 12l). Moreover, SCATS-L is a cyclic algorithm, resulting in a lower throughput on the phase serving the heavy demand. This results in long queues upstream of the entry node in both directions, especially in the non-coordinated direction. Therefore, SCATS-L has not been used to compare other performance measures. Overall, C-MP arises as the only control policy that ensures coordination in both directions along an arterial using inherent traffic properties without external constraints or fixed offsets.

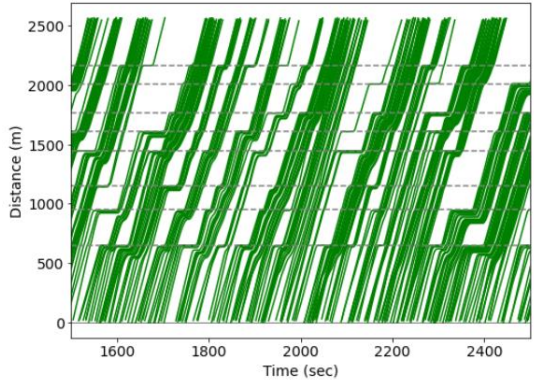
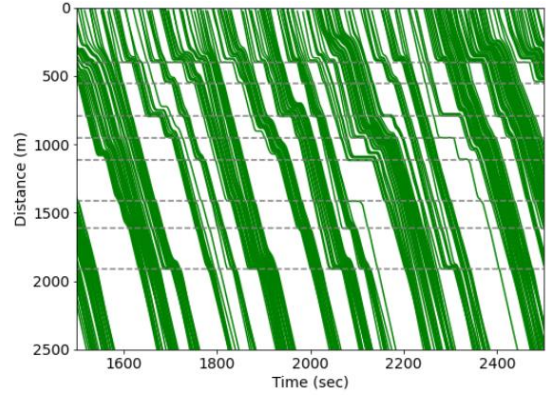

**(a) C-MP Eastbound**

**(b) C-MP Westbound**

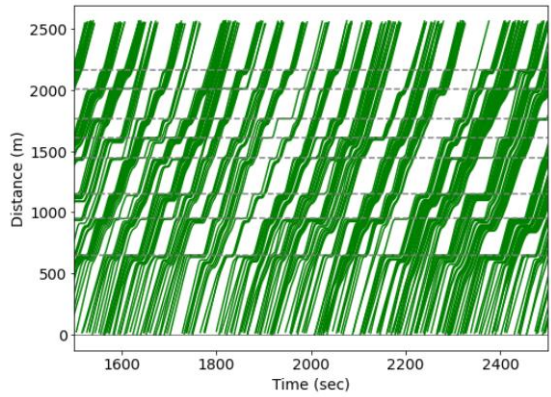
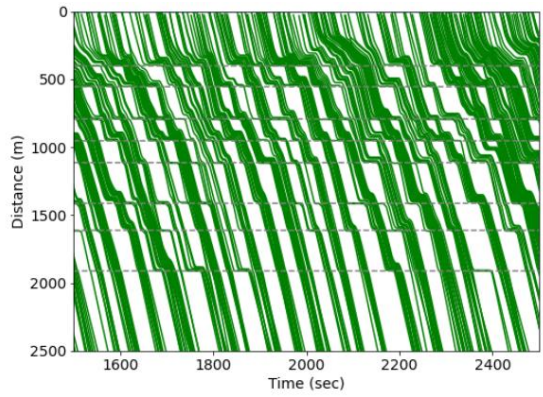

**(c) Q-MP Eastbound**

**(d) Q-MP Westbound**

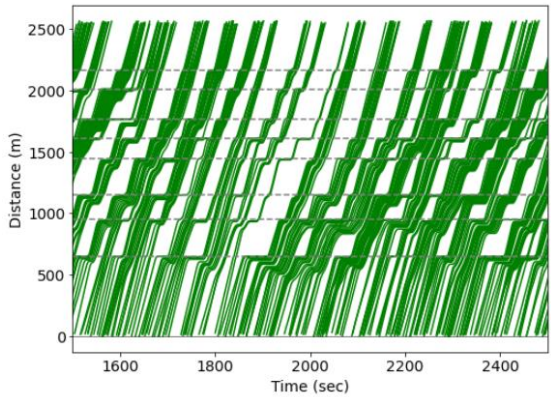
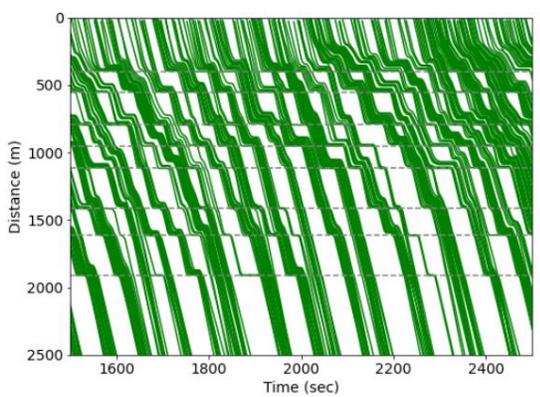

**(e) TT-MP Eastbound**

**(f) TT-MP Westbound**

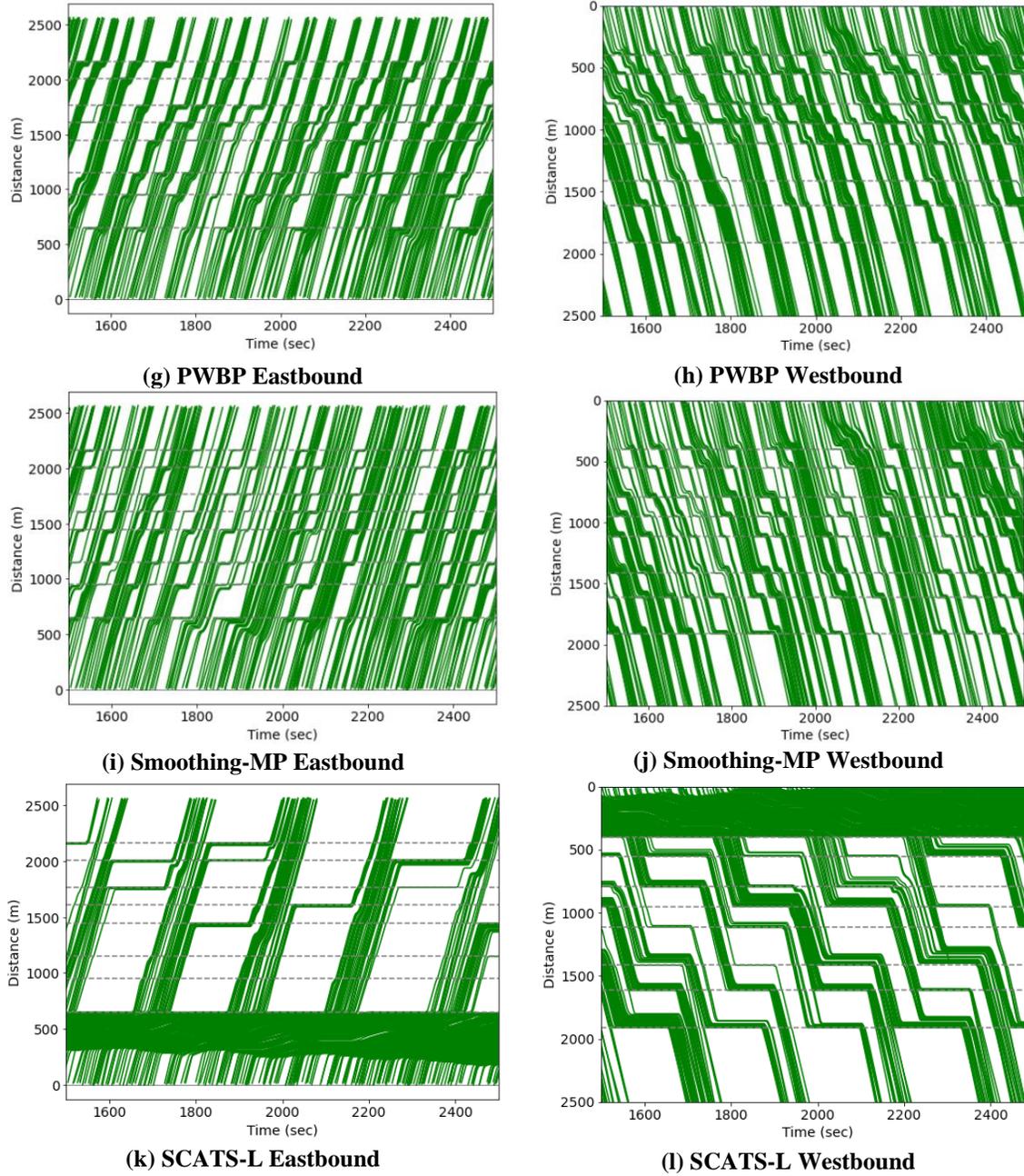

**(g) PWBP Eastbound**　　　　　　　**(h) PWBP Westbound**

**(i) Smoothing-MP Eastbound**　　　　**(j) Smoothing-MP Westbound**

**(k) SCATS-L Eastbound**　　　　　　**(l) SCATS-L Westbound**

**Figure 12. Time-space diagrams of vehicles traveling eastbound and westbound on the corridor**

### 4. 3. *Impact on fuel consumption*

One objective of a coordinated traffic signal system is to reduce the number of stops for vehicles traveling across a series of intersections which would intuitively lead to lower fuel consumption. Figure 13 provides the fuel consumption for the C-MP algorithm under various values of the tuning parameters $\alpha$ and $\beta$. Like previous figures, horizontal lines are used to denote the baseline methods. At medium demands, although the total travel time consistently increases with $\beta$ (see Figure 8a), the network fuel consumption initially drops, then starts to rise. At high demands, increasing both tuning parameters $\alpha$ and $\beta$ follow a trend similar to the change in the total travel time in the network. In both scenarios, C-MP provides the lowest fuel consumption of all methods; the most fuel efficient strategy is observed when $\alpha = 0.6$ and $\beta = 0.75$. Although PWBP corresponds to the lowest total travel time under the medium demand scenario, vehicles experience regular stops that result in increased fuel consumption compared to C-MP. Similarly, while Smoothing-MP provides similar arterial travel time performance to the C-MP under medium demands, C-MP provides better fuel efficiency by ensuring bi-directional coordination as well as balancing traffic delays in the minor directions.

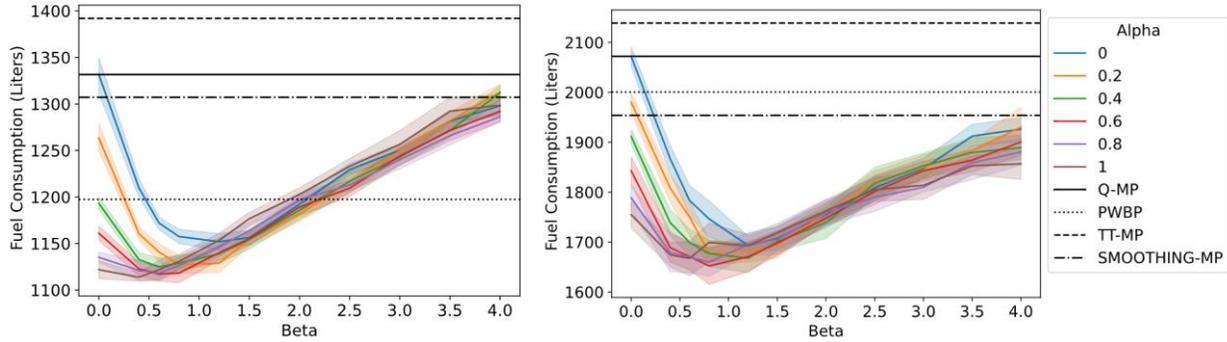

**Figure 13. Total fuel consumption: (a) medium demand; b) high demand**

Vehicles traveling along the corridor benefit from fewer stops and reduced travel time from the coordinated signal systems under the C-MP and Smoothing-MP control policies. This results in lower fuel consumption for vehicles on the corridor, as shown in Figure 14. The relationship between fuel efficiency and the C-MP tuning parameters match the trend of the corridor travel time shown in Figure 9. Increasing both $\alpha$ and $\beta$ initially leads to improved fuel economy that gradually diminish after $\beta = 1$. Additionally, this reveals that higher values of $\alpha$ and $\beta$ only lead to additional delay and subsequent idling for the vehicles on the minor direction which leads to an increase in the total fuel consumption in the network.

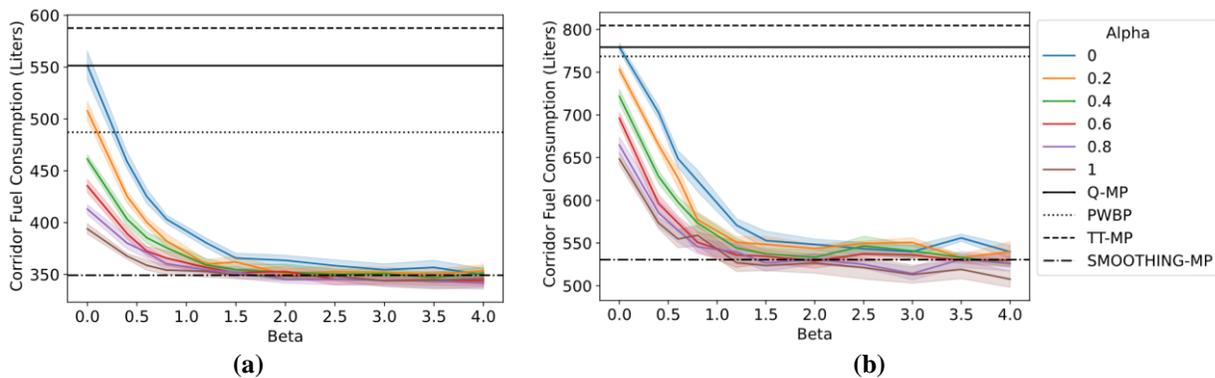

**Figure 14. Fuel consumption on corridor: (a) medium demand; b) high demand**

### 4. 4. Pareto frontiers

A trade-off exists between conflicting yet similar objectives: lowering the travel time on the corridor and lowering the total travel time. The Pareto Frontier serves as a tool to visualize and analyze this trade-off. Points on the Pareto Frontier represent optimal solutions where one objective cannot be improved without worsening the other. Points not on the frontier represent outcomes where improvements in one or more objectives are possible without a trade-off. Thus, the frontier provides a spectrum of balanced outcomes, that allow the identification of the most efficient configuration under varying traffic conditions. This sub-section uses the Pareto Frontiers to compare the performance of the proposed C-MP strategy with the baseline methods.

Figure 15 show scatter plots of the total travel time and the corridor travel time for the two demands tested. Each point indicated using round markers on the figure corresponds to the average value from 10 random seeds for a specific configuration of $\alpha$ and $\beta$ tested, and a color bar is used to indicate the sum of $\alpha$ and $\beta$ . The result of Q-MP is shown using a blue marker while the orange, purple and red markers are used to denote PWBP, TT-MP and Smoothing-MP respectively. Points that lie on the Pareto Frontier are joined using a red line. Notice that under the medium demand scenario, the Pareto Frontier comprises only the Smoothing-MP, PWBP and the C-MP for a subset of $\alpha$ and $\beta$ values. This suggests that the other methods provide worse corridor TT, total travel time, or both. For the high demand, the Pareto Frontier is entirely made up of points representing the C-MP. This suggests that the C-MP generally provides a better balance between these competing objectives than the benchmark strategies – particularly when demand is high – and the specific parameters can be used to control this tradeoff.

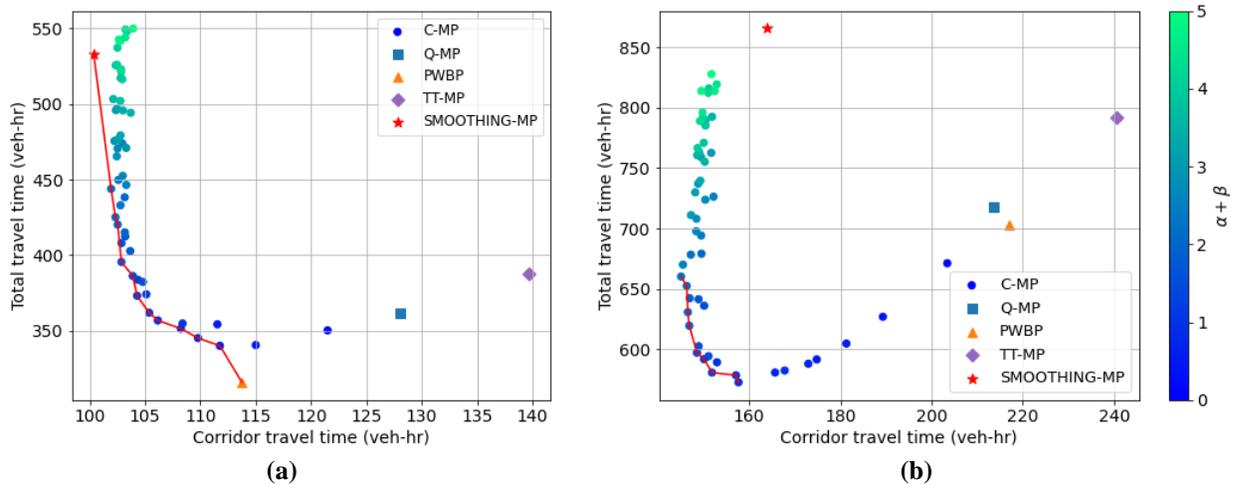

**Figure 15. Total travel time vs Corridor travel time: (a) medium demand; b) high demand**

Figure 16 provides the Pareto Frontier considering total travel time and fuel consumption. For the medium demand, the Pareto Frontier is made up of points representing the C-MP and PWBP, while the Pareto Frontier for the high demand scenario is entirely made up of points representing the C-MP.

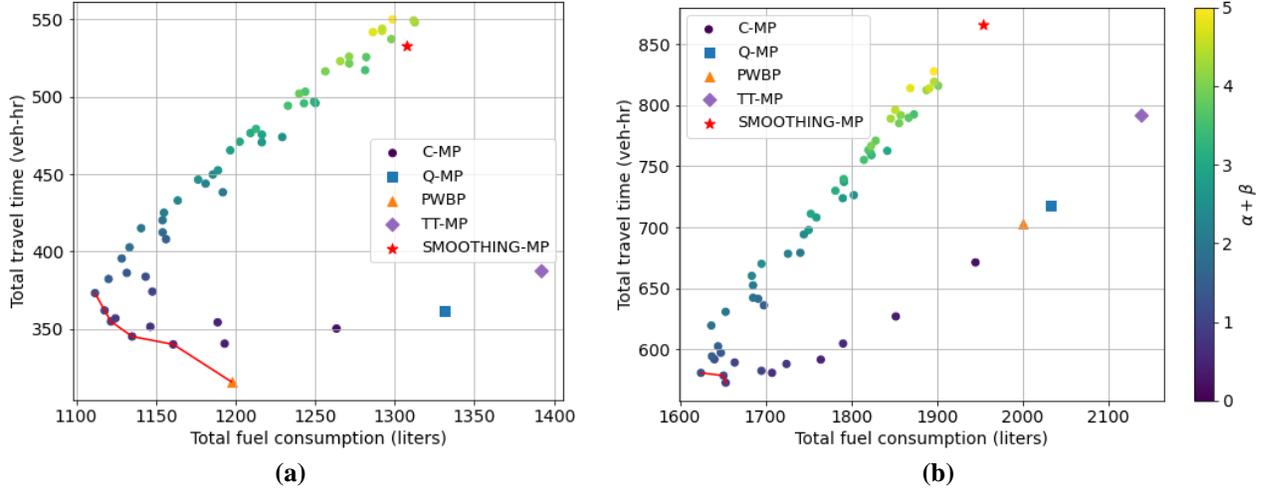

**Figure 16. Total travel time vs fuel consumption: (a) medium demand; b) high demand**

### 4. 5. Network accumulation/stability

While this study analytically proves the maximum stability property of C-MP, simulation tests were carried out to compare the stability and range of demands for which the C-MP algorithm is stable compared to the benchmark MP algorithms.

#### 4. 5. 1. Accumulation in whole network

The stable region refers to the size of the feasible demand that can be served by the control policy, i.e., the outflow of vehicles is equal to the inflow such that the number of total vehicles in the network remains bounded and does not grow over time. Different levels of demand were simulated and the evolution of vehicle accumulation in the network is shown in Figure 17 for each of the medium and high demand scenarios. The configuration of tuning parameters of C-MP selected for this analysis ($\alpha = 0.6, \beta = 1$) was selected from the knee-point of the Pareto Frontier in Figure 15.

Figure 17a exhibits a stable scenario for all MP algorithms in which the total number of vehicles in the network does not keep growing under the medium demand scenario (total vehicle entry rate of 6,336 veh/hr). The confidence intervals of C-MP and Q-MP overlap throughout the simulated period, which suggests that the arterial experiences similar accumulation levels when each of the algorithms are applied. Despite operating with a stable and non-increasing demand, the Smoothing-MP exhibits a significantly higher accumulation compared to the other benchmark methods. Under the high demand scenario (total vehicle entry rate of 7,656 veh/hr) shown in Figure 17b, the C-MP algorithm exhibits the lowest overall accumulation and is stable during the entire simulated period. However, all the other benchmark methods exhibit unstable behavior in which the accumulation of vehicles grow over time. These results not only confirm that the performance of the C-MP is stable but also reveal that the C-MP algorithm has a larger stable region than Q-MP and other benchmark algorithms.

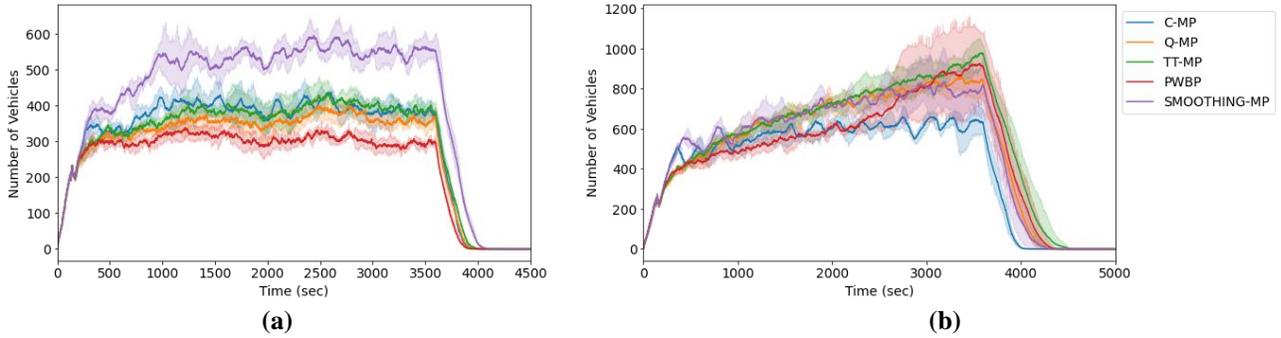

**Figure 17. Accumulation in whole network: (a) medium demand; b) high demand**

### 4. 5. 2. Accumulation on entry links in minor direction

Since C-MP prioritizes the movement on the coordinated corridor, it is expected that the vehicles entering in the minor direction may experience higher delays or longer queues. However, as demonstrated in Figure 18a-b, the accumulation of vehicles on the entry links in the minor direction are also stable and frequently served under C-MP at both demand levels. The periodic fluctuations observed in the C-MP are indicative of platooning in the minor direction that are served at less-frequent but regular intervals. While the other benchmark methods are stable at medium demand, neither is able to accommodate heavy demand evidenced by Figure 18b. Finally, Smoothing-MP coordinates the through movement in the major direction hence, results in the highest accumulation on the entry links in the minor direction among the benchmark methods. In summary, this implies that C-MP can not only prioritize the through movement in the major direction, but also serves the incoming demand on the minor direction without causing unreasonable delays.

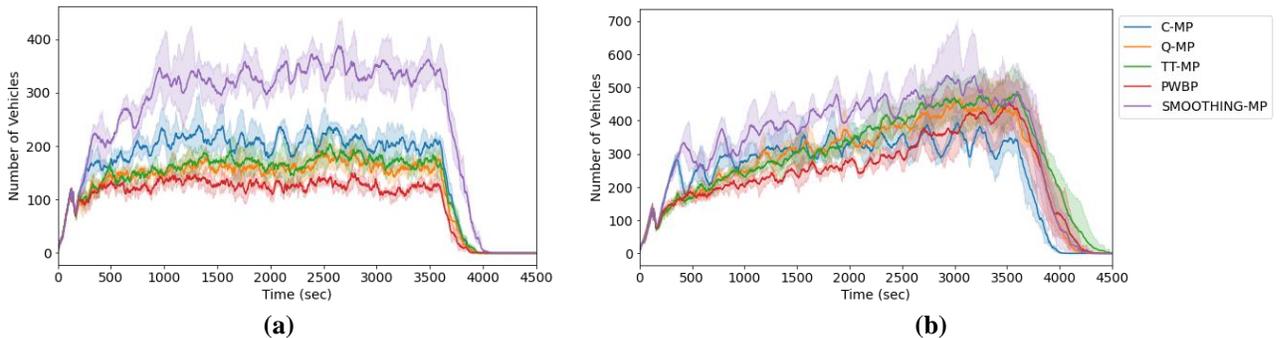

**Figure 18. Accumulation on entry links in minor direction: (a) medium demand; (b) high demand**

## 5. Conclusion

This study proposes C-MP: a computationally efficient adaptive-coordinated traffic signal control algorithm built using the max-pressure framework. The C-MP control policy utilizes both the number of vehicles and the space-mean-speed on upstream and downstream links at intersections to detect and prioritize the movement of moving platoons upstream of the signal, as well as identify space available downstream for platoons to move into. By accounting for platoons in this way, the algorithm is able to coordinate traffic signals along a corridor in both directions, allowing for more smooth traffic flow without the need for preset offsets. The strength of coordination imposed can also be controlled using a pair of tuning factors that would allow agencies the flexibility to adjust the performance across competing objectives (such as total travel time or travel time along the corridor only) according to their priorities. Furthermore, the C-MP algorithm maintains the theoretical guarantee of maximum stability in the whole network, which is a desirable property of MP-based traffic signal control algorithms.

The operational performance of the C-MP algorithm was compared to benchmark MP methods including Q-MP, PWBP, TT-MP, Smoothing-MP, and SCATS-L. The results reveal that C-MP significantly improves travel time and fuel consumption for vehicles traveling along an arterial. While coordination also leads to improvements in the travel time and fuel consumption in the entire network compared to benchmark methods, further increasing the weighing factors to moving vehicles lead to more green time allocated to the movement in the major direction and may lead to increased delays to the vehicles in the minor directions. Pareto Frontiers were also used to reveal the trade-off that exists between the total travel time and the travel time on the corridor, as well as the fuel consumption which presents directions for transportation agencies in determining the optimal configurations according to their objectives. Finally, a stability analysis further backs up the theoretical proof of maximum stability and demonstrates that C-MP has a larger stable region than the other benchmark methods meaning that it is able to accommodate more demand without queues growing indefinitely.

Although the control policy was tested on an arterial network, it can be readily applied to more complex urban networks where demand in the major direction is much higher than the demand on the cross-streets. Future work may also consider integrating transit signal priority with the C-MP to see if the presence of transit affects coordination in an arterial.

**ACKNOWLEDGEMENTS**
This research was supported by NSF Grant CMMI-1749200.

**AUTHOR CONTRIBUTIONS**
The authors confirm contribution to the paper as follows: study conception and design: TA, HL, VG; analysis and interpretation of results: TA, HL, VG; draft manuscript preparation: TA, HL, VG. All authors reviewed the results and approved the final version of the manuscript.